# Crystallography under external electric field

**Semen Gorfman**[a]*, **Oleg Schmidt**[a], **Vladimir Tsirelson**[b], **Michael Ziolkowski**[a], **Ullrich Pietsch**[a]

Structural response of crystals to an applied external perturbation is important as a key for understanding microscopic origin of physical properties. Experimental investigation of structural response is a great challenge for modern structure analysis. We demonstrate how advanced X-ray diffraction techniques facilitate probing tiny ($10^{-4}$ Å) distortions of bond lengths under a permanent electric field. We also discuss details of the experimental procedure essential for reaching such precision.

We ask whether the experiment can be used to evaluate chemical bonds in crystals by their sensitivity to an external electric field and discuss if the bond deformations can be predicted using the bond-valence model or the Bader's theory of atoms in molecules and crystals. Finally, we describe the new time-resolved studies of a structural response to a dynamical switch of applied electric field. These results give access to the time-lining of piezoelectric effect on a microsecond time scale.

* Corresponding Author
 Fax: 0049 271 740 3763
 E-Mail: gorfman@physik.uni-siegen.de
[a] Department of Physics
 University of Siegen
 Walter-Flex str. 3, 57072, Siegen
[b] Quantum Chemistry Department,
 Mendeleev University of Chemical Technology,
 Moscow, Russia

## Introduction

X-ray diffraction is the standard route to the determination of space group symmetry, atomic positions, electron density and thermal displacement parameters in crystals. The instrumentation is widely available in the form of commercial diffractometers incorporating the well-established theories, algorithms, and techniques to study atomic structure of crystals under static ambient conditions.

There is rapidly increasing interest in the investigation of a crystal structural response caused by the change of external conditions such as electric field, temperature, stress, short laser pulses, etc. The instrumentation for that is still quite elaborate and much less common than for ordinary X-ray diffraction. Therefore, current efforts strongly focus on the development of the experimental techniques, theoretical background, strategies for data collection, and methods for data analysis for probing small structural variations under external fields.

The interest in structural response is inspired by both fundamental and material science since the sensitivity of a structure to an external perturbation defines its physical or chemical properties. For example, direct piezoelectric effect is the appearance of a dielectric polarization in response to an applied mechanical stress; converse piezoelectric effect is the development of a macroscopic strain in response to an applied electric field, etc. [1],[2]. These properties are well understood at the macroscopic level. It is however much less clear how they are linked to atomic structure and interatomic interactions.

The role of the interatomic interaction can be illustrated with an example of $\alpha$-SiO$_2$ (quartz) and $\alpha$-GaPO$_4$ (gallium phosphate) piezoelectric materials. Both crystals belong to the same structural type (space group P3$_1$21): they are formed by a chain of corner-shared AO$_4$ tetrahedra (A = Si, Ga, P); the $\alpha$-SiO$_2$ structure converts into the $\alpha$-GaPO$_4$ by alternating replacements of a Si atom by Ga and P atoms in the tetrahedra. This replacement of the atoms doubles the magnitude of the piezoelectric constant: $d_{11}$ = 2.2 pC/N for $\alpha$-SiO$_2$, $d_{11}$ = 4.4 pC/N for $\alpha$-GaPO$_4$. Thus, the macroscopic effect originates from the change in the atomic sizes and the atomic interactions.

Understanding how and why substitution of atoms in a structure modifies its physical properties requires knowledge of material response at the microscopic rather than the macroscopic level. One must be able to predict the response of individual structural units – bond lengths, bond angles and structural polyhedra – to the different kinds of perturbation. Following from this, physical properties of a material could be evaluated even before it is synthesized, or the role of a certain chemical element in a structure could be explained. Unfortunately, the *ab initio* approach to this problem does not exist yet. Brown [3],[4],[5],[6] developed the empirical bond-valence model which allows of the calculation of the effective bond force constants and estimates the bond compressions under applied pressure using the bond lengths and the bond-valence indices. It is, however, unknown how well the bond-valence method performs when another perturbation, e.g. an external electric field, is applied.

This work aims to summarize our recent efforts and achievements in X-ray crystallography under an applied external electric field performed on inorganic single crystals. We describe the modern experimental techniques and specify typical pitfalls and technical challenges. We pay particular attention to the chemical aspect of the work and demonstrate how X-ray diffraction by crystals under an external electric field can be applied to characterize chemical bonds by their sensitivity/response to an external electric perturbation. We also discuss the ability of the bond-valence model and Bader's theory of atoms in molecules and crystals to predict the microscopic response under an applied electric field. Finally, we present a novel experimental approach to time-resolved X-ray diffraction studies of a crystal response to an electric field on the microsecond timescale and discuss the time regimes of the microscopic and macroscopic deformations.

**Experimental technique**

The magnitude of an external electric field that can be applied to an inorganic piezoelectric single crystal ranges between ~ $10^6$ and $10^7$ V/m; further field increase usually leads to an electrical breakthrough. It is smaller by a factor of $10^{-5} - 10^{-4}$ than the average internal electric field: the structural shifts induced by an external electric field are expected to be very small. Therefore, detection of the related changes in diffraction intensity is an experimental and technical challenge.

We constructed a dedicated data acquisition system that allows us to detect tiny (~0.5 %) differences of diffraction intensity under an applied electric field. The concept of the system originates from the works of Puget & Godefroy[7] and Fujimoto [8] – it was extensively used in a number of studies ([9], [10], [11], [12], [13], [15], [16], [18] [22], [21], [23],[24],[25],[30],[31],[32],[33],[34],[35]). This data acquisition strategy exploits the stroboscopic approach: an applied electric field is periodically modulated in a 4-step mode (the modulation periods consist of successive positive, zero, negative and zero states) with the frequency of about 20 Hz; signals coming from an X-ray photon counter are continuously distributed over four counting channels, synchronized with the above modulation of the high voltage. In this way, the diffraction intensity is measured quasi-simultaneously for each of the different electric field states.

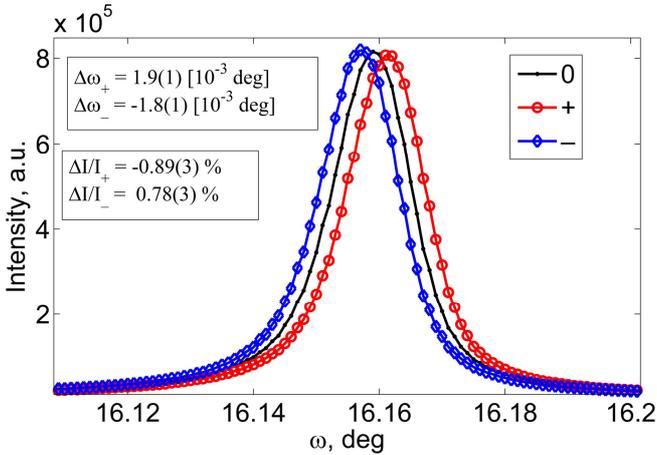

Figure 1. Typical modification of a Bragg rocking curve under applied external electric field. The right and left curves correspond to the positive and negative polarities of applied voltage, the middle curve corresponds to the average between two zero states. Shifts of the angular positions and relative change of intensity are extracted from the curves and analysed independently from each other.

The stroboscopic data collection can be applied to a variety of diffraction geometries (single crystal, powder, grazing incidence, triple axis, etc.) where diffraction intensity is measured with a point detector. For a single crystal diffraction experiment we implement an open point detector and collect rocking curves around the pre-selected Bragg positions (see [23] for schematics of the scans in a reciprocal space). An example output (the rocking curves collected stroboscopically at the positive (+), negative (−) and zero applied voltages) is shown in Figure 1.

The first reason to apply the stroboscopic technique is to avoid any irreversible changes caused by a perturbation. The intervals of a quasi-static electric field must be long enough to move the atoms within a unit cell and to change the lattice dimensions. On the other hand, they should be short enough to prevent any mesoscopic changes such as migration of free charges, variation of defects structure, etc. These mesoscopic changes are difficult to model; their contribution to the total X-ray diffraction intensity can be higher than that of the microscopic structural changes (shifts of atomic positions in a unit cell). Thus, the modulation period of an applied high voltage should be chosen according to the expected time dynamics of the processes induced by an electric field.

The second reason to apply the stroboscopic technique originates from the extremely small intensity variations. As a rule, Bragg intensities change by less than 1 %. All "external" errors such as instability of a synchrotron beam, drift of sample temperature, or variations of detector efficiency must be carefully excluded. For that reason, the diffraction profiles corresponding to the different states of an electric field must be collected quasi-simultaneously, i.e. stroboscopically.

**Comments on data analysis**

Besides using the stroboscopic experimental technique, we need to analyse the data differently from that of conventional X-ray structure analysis.

In particular, we perform a structural least-square refinement against the relative intensities variation, not the absolute values of intensities. The relative intensity variation is modelled as

$$\left(\frac{\Delta I}{I}\right)_{MOD} = \frac{|F(E)|^2 - |F(0)|^2}{|F(0)|^2}, \quad (1)$$

where $|F(E)|^2$ is the structure factor under applied electric field $E$. Theoretical analysis [19],[20] has shown that the field-induced displacements of atoms give the only observable contribution (more than 0.1 %) to the change of kinematical diffraction intensity. Therefore, they are used as the variables in refinement.

The relative change in intensity is extracted from the experimental data (Figure 1):

$$\left(\frac{\Delta I_\pm}{I_0}\right)_{EXP} = \frac{I_\pm - I_0}{I_0}, \quad (2)$$

where $I_+, I_-, I_0$ are the integrated intensities (areas under the rocking curves) corresponding to the positive, negative and zero (average of two zeros) states of an electric field correspondingly.

The approach using the relative intensity variations instead of the absolute intensity has the following advantages. First, it benefits from cancelation of several sources of systematic errors such as absorption, polarization correction, scaling of different data sets, etc. It is particularly important as *in situ* electrical contacting of a crystal introduces several absorption related corrections to the absolute diffraction intensity and further increases the systematic error. Second, traditional techniques of structure analysis would fail to distinguish between field-on and field-off structures differing by ~ $10^{-4}$ Å atomic displacements. The scheme that uses the intensity variations allows one to obtain the field-induced displacements of atoms directly, without referring to their precise initial positions.

To demonstrate that we suppose that the initial position of an atom is $R_0 (\pm \delta R)$, where $\delta R$ is the systematic error of the $R_0$, and $\Delta R$ is an electric field induced displacement of the same atom. Expanding the electric field-induced variation of intensity in a second-order Taylor series we obtain:

$$\Delta I = I(R_0 + \delta R + \Delta R) - I(R_0 + \delta R) =$$
$$= \frac{\partial I}{\partial R} \Delta R + \frac{\partial^2 I}{\partial R^2} \delta R \cdot \Delta R + \frac{1}{2} \frac{\partial^2 I}{\partial R^2} \Delta R^2 \quad . \quad (3)$$

The systematic error, $\delta R$, contributes to the intensity changes as a second order term only. To demonstrate how it affects the relative changes in intensity, we simulated $\Delta I/I$ values in the case of the $Li_2SO_4 \cdot H_2O$ crystal for a few random $\delta R$ sets. Both initial atomic positions and the small (~$10^{-4}$ Å) electric field induced displacements from Schmidt et al. ([25]) have been used. We randomly shifted the field-free atomic positions by $|\delta R| = 10^{-3}$ Å (one order of magnitude higher than the field induced displacements) and calculated (using Eq. 1) the intensity changes due to additional electric field induced shifts. Figure 2 demonstrates the results – electric field induced intensity variations of three selected Bragg peaks for 50 different sets of $\delta R$.

Figure 2. Relative variations of the structure factors of $Li_2SO_4 \cdot H_2O$ calculated for 50 different sets of initial atomic positions. Each point corresponds to the field-free position of each atom in a unit cell displaced in a random direction by $10^{-3}$ Å. The relative intensity variation (in %) was calculated using Eq. (1) and electric field induced displacements from [25].

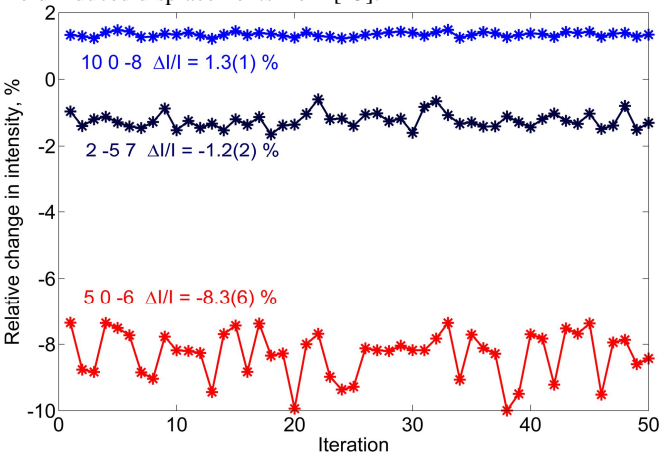

This modelling reveals how the systematic error, $\delta R$, in the initial atomic position, propagates to the systematic error in the relative variation of structure factors. It shows that consideration of intensity variations instead of absolute intensities relaxes the requirements of precise knowledge of the field-free structure: it is acceptable to determine the initial atomic positions with the precision of $10^{-3}$ Å. In this case, the systematic errors of modelled intensity variations are still smaller than the average systematic error of measured intensity variations.

**The problem of extinction**

Extinction can create serious problems for the data analysis. It accounts for the correction factor *y* to the kinematical X-ray diffraction intensity in the case when the diffraction is not entirely kinematical, $I = y|F|^2$. Although a few approximate models are available to estimate extinction correction (see the review in [45]), they are never accurate enough to consider extremely small changes of intensity. In the case of a plate-shaped perfect crystal Hansen et al [17] suggested to use the alternative expression $I \sim |F|^n$ where *n* ($\neq 2$) approximates the precise solution given by the dynamical X-ray diffraction theory ([50]). For a real crystal, i.e. a crystal with an uncertain degree of perfection, the so-called primary and secondary extinction corrections must be introduced and modelled as the functions of the mesoscopic structural parameters [51] – the average size of a hypothetical mosaic block and the distribution of mosaic block orientations. Any variation of a mosaic structure reflects in an extinction correction. Unfortunately, neither the changes of a mosaic structure in a crystal under an external electric field nor the related changes of extinction correction can be simulated and introduced into the refinement procedure. As a result, the Bragg reflection, presumably affected by extinction, must be excluded from the refinement.

We adopted the following guidelines to avoid the Bragg reflections strongly affected by extinction: a) reflections with lower values of structure factors are measured; the extinction correction is expected to be significant for strong Bragg peaks only, b) sets of symmetry equivalent reflections are collected; the measured intensity changes are compared – significant disagreement indicates the appearance of an extinction, c) the change of a peak width (FWHM) is controlled: those peaks with significant change of the FWHM under applied electric field are not included in the refinement. The application of the selection rules b) and c) rejects approximately 30 % of the reflection (the exact percentage of rejected reflections strongly depends on the crystal). The inclusion of these reflections in the refinement results in a dramatic increase of the reliability factors and unphysical (extraordinarily large) refined values of atomic displacements.

The alternative approach to the problem of extinction relies on the different time dependences of the structural and mesoscopic responses: the atomic shifts are much faster than the change of extinction. Therefore, time resolved measurements of a rocking curve can be a potential way to separate different contributions to the intensity. The new experimental technique which can be applied in future is described in the last part of this paper.

## Macroscopic strains and piezoelectric coefficients

Shifts of the peak positions (Figure 1) can provide information about a field induced macroscopic strain. The peak shifts

$$(\Delta\omega_{\pm})_{EXP} = \langle\omega_{\pm}\rangle - \langle\omega_0\rangle, \quad (4)$$

with average positions of a rocking curve given by

$$\langle\omega_{\pm,0}\rangle = \frac{\int_{-\infty}^{+\infty} \omega I_{\pm,0}(\omega)d\omega}{\int_{-\infty}^{+\infty} I_{\pm,0}(\omega)d\omega},$$

are analysed using the approach introduced by Graafsma [27] and further developed in [23]. The observed peak shifts are modelled as:

$$(\Delta\omega)_{MOD} = -tan\theta \cdot \frac{d_{ijk}E_iH_jH_k}{H^2} - \frac{d_{ijk}E_iY_jH_k}{H} + \frac{R_{ijk}E_iY_jH_k}{H}, \quad (5)$$

where $\theta$ is the Bragg angle, $H$ is the reciprocal lattice vector length, $H_i$ and $d_{ijk}$ are the components of the reciprocal lattice vector and the piezoelectric tensor correspondingly (the components are given relative to the crystal physical Cartesian coordinate system). The unit length vector $Y$ is the cross product $Y = [H,w]/H$, where the unit vector $w$ points along the rotation axis of the diffractometer. The second rank tensor $R_{ijk}E_i$ describes the rotation of the whole crystal under an applied electric field.

Table 1. The piezoelectric coefficients ($10^{-12}$ m/V) of $Li_2SO_4 \cdot H_2O$ from the analysis of 20 Bragg peak shifts ([25]), in comparison to the piezoelectric coefficients determined in a dynamical pressure cell ([36],[37]).

|       | $d_{211}$ | $d_{222}$ | $d_{233}$ | $d_{213}$ |
|-------|-----------|-----------|-----------|-----------|
| X-ray | -3.4(3)   | 15.3(2)   | 1.4(3)    | -2.4(2)   |
| DPC   | -3.3(2)   | 15.8(5)   | 1.7(1)    | -2.2(1)   |

We successfully apply this algorithm to different crystals, including crystals of low symmetry ([23], [25], [38]) and routinely use it for determination of piezoelectric coefficients. As an example, Table 1 compares the piezoelectric constants of monoclinic $Li_2SO_4 \cdot H_2O$, determined by using the above algorithm, with those obtained by the application of a dynamical pressure cell ([36],[37]). It demonstrates that the measured peak shifts are a reliable probe of a piezoelectric strain induced by an external electric field.

In conclusion, a single rocking curve measured stroboscopically under an electric field gives an independent probe of both macroscopic and microscopic structural variations measured under the same conditions: spot of the sample, temperature, etc.

## Distortion of chemical bonds under external electric field

Let us consider the microscopic structural changes in the two investigated systems: $\alpha$-$GaPO_4$ and $Li_2SO_4 \cdot H_2O$ crystals.

The $\alpha$-$GaPO_4$ structure can be viewed as a chain of alternating corner-sharing $GaO_4$ and $PO_4$ tetrahedra (Figure 3, right). The Ga–O/P–O chemical bonds have different length, strength and type. In addition, Ga and P cations are characterized by significantly different formal atomic charges.

The crystal structure of $Li_2SO_4 \cdot H_2O$ is formed by $LiO_4$, $LiO_3(H_2O)$ and $SO_4$ groups linked together by O atoms to form a three dimensional tetrahedral framework, see Figure 3, left.

As a first step we studied the static properties of Ga–O, P–O, Li–O and S–O chemical bonds using two different approaches: the bond-valence theory and the theory of atoms in molecules and crystals.

1. The empirical bond-valence model developed by Brown *et al.* ([4],[5],[6]) estimates effective bond force constants, considering that bond distances and bond valences are known.

2. The theory of atoms in molecules and crystals developed by Bader [39] quantifies chemical bonds using experimental or theoretical electron densities. Interatomic interactions are rated by the features of bond critical points (saddle points in the electron density on bond paths – lines of locally maximal electron density linking some atomic pairs within the structure). These features are the electron density values, $\rho_{BCP}$, Laplacian of electron density, $\nabla^2\rho_{BCP}$, and the total electronic energy density, $h_{BCP}$. The energy density is computed from the electron density and its derivatives using Kirzhnits approximation ([41],[42]). According to [43],[44],[45],[46],[47],[48], and [49], these properties categorize each bond as either a shared (covalent), a closed-shell (ionic) or an intermediate interaction. The number of electrons assigned to each atom can be evaluated by integrating the electron density within an atomic basin (volumes inside zero flux surfaces of the electron-density gradient).

We applied this approach to the electron densities of $\alpha$-$GaPO_4$ and $Li_2SO_4 \cdot H_2O$ derived from the X-ray diffraction experiments and using Hansen & Coppens multipole model ([28],[45],[46]). The data were collected at the D3 beamline at HASYLAB (for $\alpha$-$GaPO_4$) and home-lab CAD4 diffractometer (for $Li_2SO_4 \cdot H_2O$). The details of the refinement procedure are described in [21] (for $\alpha$-$GaPO_4$) and [25],[38] (for $Li_2SO_4 \cdot H_2O$). All the experimental results, including the features of electron density at the bond critical points, were validated by density functional calculations performed with the WIEN2k package [29] (with the exception of Laplacian of electron density at the S-O critical point, where different signs of $\nabla^2\rho_{BCP}$ were obtained from the experimental and theoretical electron density). Further details of the DFT calculations can be found in [21] (for $\alpha$-$GaPO_4$) and [25],[38] (for $Li_2SO_4 \cdot H_2O$).

Table 2. Properties of Ga–O, P–O (α-GaPO$_4$) and Li–O, S–O (Li$_2$SO$_4$·H$_2$O) chemical bonds, including a) bond force constant, K, according to the bond-valence model ([3]); b) cation charge according to Bader's theory of atoms in molecules; c) electron density at the bond critical point; d) Laplacian of the electron density at the bond critical point; e) total electronic energy density at the bond critical point (in atomic units); f) type of interactions according to topological properties of the bonds (see the references in the text) where C, I and CI denote covalent (shared), ionic (closed shells) and intermediate atomic interactions, respectively. All properties b) - f) are based on the experimental electron densities. For S-O bond only we obtained different values of the Laplacian from the experimental and the theoretical electron densities, both are presented in the table (experimental/theoretical).

|  | Ga–O | P–O | Li–O | S–O |
|---|---|---|---|---|
| a) K, $N/m$ | 313 | 833 | 55 | 1130 |
| b) Q, $e$ | 1.19 | 3.13 | 0.90 | 4.10 |
| c) $\langle\rho_{BCP}\rangle$, $e/Å^3$ | 0.79 | 1.70 | 0.16 | 1.90 |
| d) $\langle\nabla^2\rho_{BCP}\rangle$, $e/Å^5$ | 13.73 | 13.49 | 4.87 | 8.6 / -2.8 |
| e) h, a.u. | -0.03 | -0.24 | 0.01 | -0.31 |
| f) Interaction type | CI | CI | I | CI / C |

Table 3. Summary of the Ga–O, P–O, Li–O and S–O bond response to an applied electric field: a) number of collected Bragg peaks; b) number of model parameters used for the refinement of atomic displacements; c) average value of intensity variation; d) reliability factors, e) average deformation of chemical bond under normalized (E = 1 kV/mm) electric field; f) bond distortions, roughly estimated using the bond-valence theory and calculated as $2Q·E/K$, where Q is the Bader's charge of the cation, $E = 1$ kV/mm, K is the bond force constant from Table 2.

|  | Ga-O | P-O | Li-O | S-O |
|---|---|---|---|---|
| a) Reflections | 54 |  | 130 |  |
| b) Parameters | 16 |  | 13 |  |
| c) $\langle\Delta I/I\rangle$, % | 0.8 |  | 1.6 |  |
| d) $R_w / R$ | 0.21 / 0.20 |  | 0.23 / 0.29 |  |
| e) $\langle\Delta d\rangle$, $10^{-5}$ Å | 1.8 | 4.1 | 11.3 | 1.6 |
| f) $2Q·E/K$, $10^{-5}$ Å | 1.2 | 1.2 | 5.2 | 1.2 |

The bond-valence model yields force constants of the Ga–O, P–O, Li–O and S–O interactions, 313, 833, 55, 1130 N/m correspondingly. Properties of the Ga–O, P–O, Li–O, and S–O bonds based on topological analysis are summarised in Table 2. It shows that the Li–O interactions are ranked as closed-shell (ionic) ones (both the Laplacian of the electron density and the total electronic energy density is positive). The S–O bond is rated as either covalent (as follows from the theoretical electron density) or intermediate (as follows from the experimental electron density) between closed-shell and covalent. The bonds in α-GaPO$_4$ are intermediate between ionic and covalent (we use the terminology given in [39] and categorize chemical bonds by the signs of Laplacian of electron density and electronic energy density).

The features of the chemical bonds as reflected by the experimental electron densities (the results of the multipole refinements) are shown in Figure 4. They are plotted in four sections (covering O–X–O planes, X = Ga, P, Li, S) in the form of the static deformation electron density: the difference between multipole model electron density (the density, based on the refined multipole model parameters) and a sum of the densities of hypothetical independent spherical atoms located at the same positions as atoms in a crystal. The difference density maps characterize the properties of bonds qualitatively since they show explicitly the redistribution of the density into the bonding areas between the atoms.

As the next step, we analysed the chemical bonds by their ability to respond to the electric field, applied along the [110] direction for α-GaPO$_4$ and the [010] direction for Li$_2$SO$_4$·H$_2$O. Table 3 summarizes selected details of the X-ray diffraction experiments and the model refinement: the number of measured reflections, the number of parameters of the model, average intensity variation and reliability factors. The choice of free model parameters, constrains due to the symmetry and pseudo-symmetry of the atomic positions, constrains due to the fixed dielectric polarization, etc. were described elsewhere ([20],[21],[22],[25],[38]). The relative standard uncertainties of the bond deformation were estimated at less than 1 % for both crystals. The last two rows of Table 3 show the distortions of the chemical bonds reduced to E = 1 kV/mm, both calculated from the refined displacements of atoms and estimated according to the bond-valence theory.

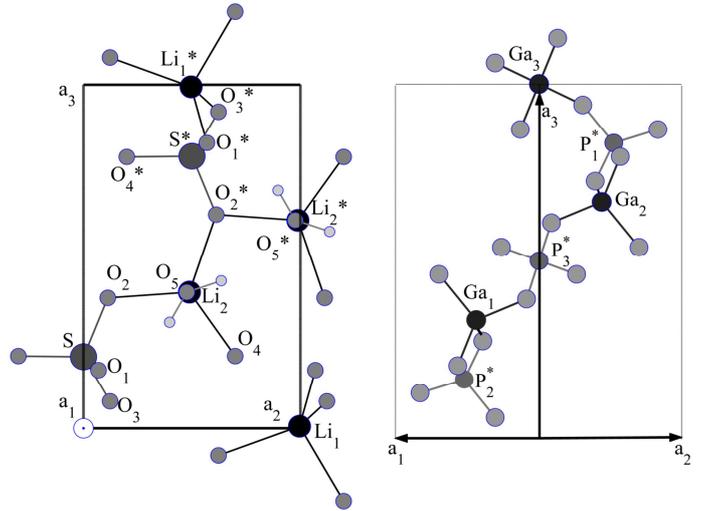

Figure 3. The crystals structures of Li$_2$SO$_4$·H$_2$O and α-GaPO$_4$, studied under permanent external electric field. Left: crystal structure of Li$_2$SO$_4$·H$_2$O in projection along [100] direction, the atoms related by a symmetry operation are marked by stars. Right: α-GaPO$_4$ structure is shown in projection along [110] direction.

The example of the atomic displacements in the PO$_2$ fragments of α-GaPO$_4$ is presented in Figure 5.

Figure 4. Experimental deformation electron densities (the difference between experimental density of a crystal and density of spherical atoms placed in the same points ([45])), in the O–Li–
5

O, O–S–O and O–Ga–O, O–P–O planes. Contour spacing is 0.05 e/Å$^3$, negative (density deficient) contours are represented by the broken lines. The features of the density accumulation in the interatomic regions approximately characterize the type of the bonding.

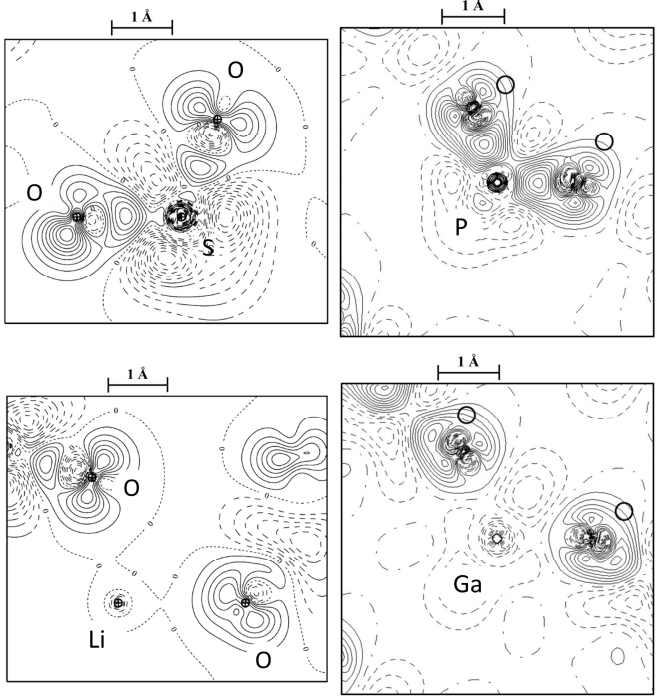

Figure 6 presents the magnitudes of the bond-length distortions (reduced to E=1 kV/mm) as a function of cosα, where α is the angle between the electric field and the bond line.

Figure 5. The schematic view of the atomic displacements in PO$_2$ sublattice of α-GaPO$_4$. The directions of atomic displacements induced by the electric field applied along [110] are shown by arrows.

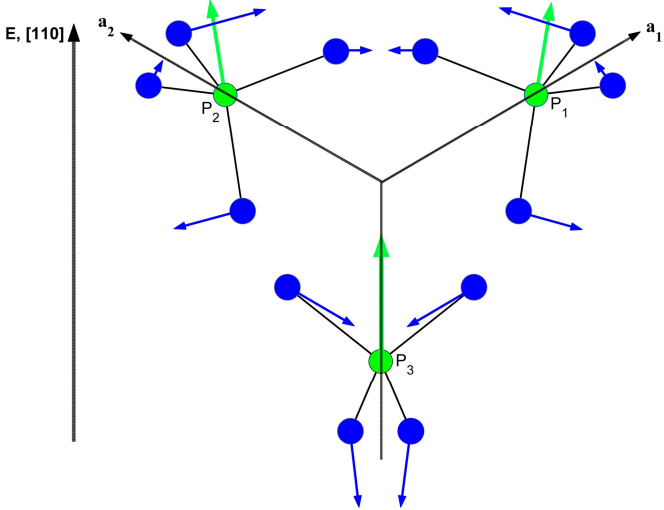

The correlation between cosα and Δd is observed for the Li–O bonds only. In addition, Li–O shows the highest response to an electric field: their average distortion is 11.3·10$^{-5}$ Å/(kV/mm), referring to the ionic character and low strength of the Li–O interaction. The S–O bonds are the most resistant to an applied external electric field: their distortions average to 1.6·10$^{-5}$ Å/(kV/mm). It agrees with the strength of the S–O bonds: the largest force constant, electron density and electronic energy density at the bond critical point. This situation is different in α-GaPO$_4$: the stronger P–O bonds are more responsive to an electric field than the weaker Ga–O bonds: the bond distortions average to 4.1·10$^{-5}$ and 1.8·10$^{-5}$ Å/(kV/mm) correspondingly. This contradiction may account for the different P and Ga atomic charges: the force exerted on the bond is formally proportional to the charges of the bounded atoms. Unfortunately, the interpretation based on the atomic charges is valid for isolated charges or ions only. The picture becomes more complicated if the atoms are involved in shared interactions where significant charge density accumulates in the interatomic region. For example, the atomic charge treatment is not applicable to the S–O interactions: the atomic charge of S atom is high (+4.1 e) but the bond distortion is low. Thus, the analysis of the experimentally derived small S–O bond distortions needs more sophisticated treatment models.

Figure 6. Distortion of Ga–O, P–O, Li–O and S–O chemical bonds under external electric field E = 1 kV/mm, applied along [110] (for α-GaPO$_4$) and [010] (for Li$_2$SO$_4$·H$_2$O) direction. Distortions of covalent S–O, Ga–O, P–O are strongly anisotropic and not obviously correlated with the bond-field angle. Distortion of ionic Li–O bonds is linearly dependent on the projection of electric field in the bond direction.

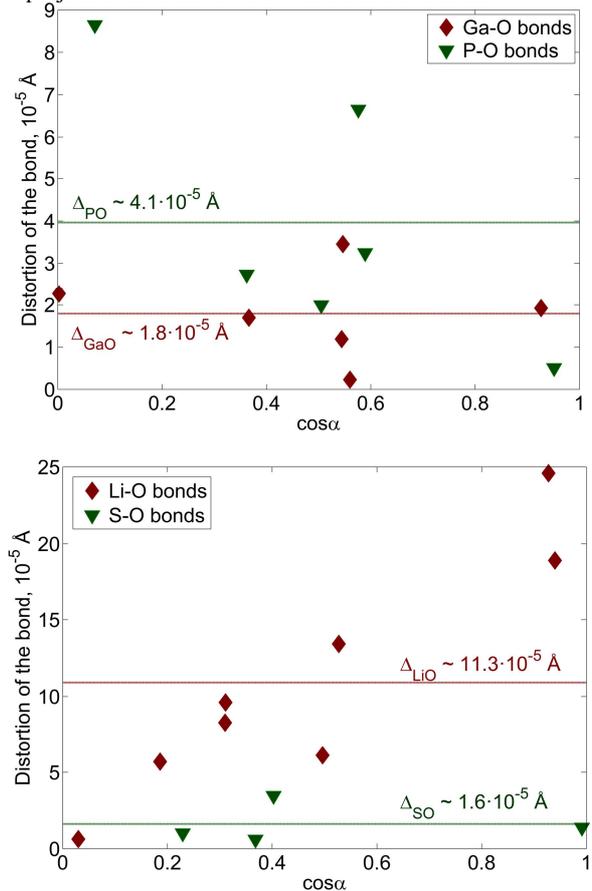

## Time-resolved studies of piezoelectric crystals

X-ray diffraction under quasi-static electric field is a tool to probe atomic shifts and macroscopic deformation simultaneously and independently from each other. To understand the interplay between microscopic and macroscopic shifts we need to compare their time regimes, i.e. perform the experiment in a time-resolved mode. The fundamental question is whether (1) microscopic shifts (changes of bond length) create macroscopic deformation or - *vice versa* - (2) elastic deformation (changes of lattice dimensions) creates a microscopic strain. Alternatively (3) microscopic and macroscopic deformation can be in the permanent interplay with each other so that their dynamics is not separable. In the first scenario, the piezoelectric deformation starts with the shifts of atoms in a unit cell and finishes with the development of a macroscopic deformation, releasing the residual internal stress of the distorted bonds. In the second scenario, an electric field induces a macroscopic deformation first; it results in the shifts of atoms – they are pushed into the positions that are optimized for the new unit cell size. The third scenario can be described as a mixture between the first two: electric field shifts the atoms in the unit cell first; the macroscopic strain develop later; the atoms are pushed further due to the macroscopic deformation, etc., so that the process goes on until the self-consistency between the atomic positions, the electric field and the lattice constants is achieved.

The first X-ray diffraction study of the relaxation processes in piezoelectric crystals has been performed by van Reeuwijk *et al.* ([14]). The pump-probe based technique and the special time structure of the beam has been used.

In this work we developed the novel data acquisition system that facilitates the time-resolved X-ray diffraction study of a single crystal under a periodic electric field and without using time structure of a synchrotron beam. While the high-voltage is cycled, the data acquisition system time-stamps the detected photons relatively to the beginning of a high-voltage cycle and distributes them in different time channels (Figure 7). The width of a single channel is 100 ns, the typical period of applied perturbation is 1 ms, giving 10000 rocking curves, at the end of the experiment. Binning of the neighbouring channels can improve counting statistics; the time resolution changes accordingly. The main advantage of the proposed scheme is the possibilities to read out all time channels simultaneously and perform the time-resolved measurements without using the time structure of the synchrotron radiation. Thus, the data collection is much faster than in the commonly used pump-probe method.

In addition, we designed the electronic circuits supplying dynamic switches of the field: the rising of 3 kV of voltage during 200 ns.

Figure 7. The flow chart of the data acquisition system for the time-resolved X-ray diffraction study of crystals under dynamically switched electric field. The high-voltage generator (upper left corner) produces the periodic high-voltage signal: a single period includes long intervals of opposite polarities (half of a period each) and dynamic (3 kV/200 ns) switches between them. The digital pulse generator (middle left) controls the frequency of the high-voltage: each digital pulse corresponds to the beginning of a new high voltage cycle. The high-voltage is delivered to the crystal. Both, digital pulses and detector pulses are introduced into the FPGA-board (Field-programmable gate array). FPGA processes the detector signals and time-stamps them relatively to the beginning of the high-voltage cycle.

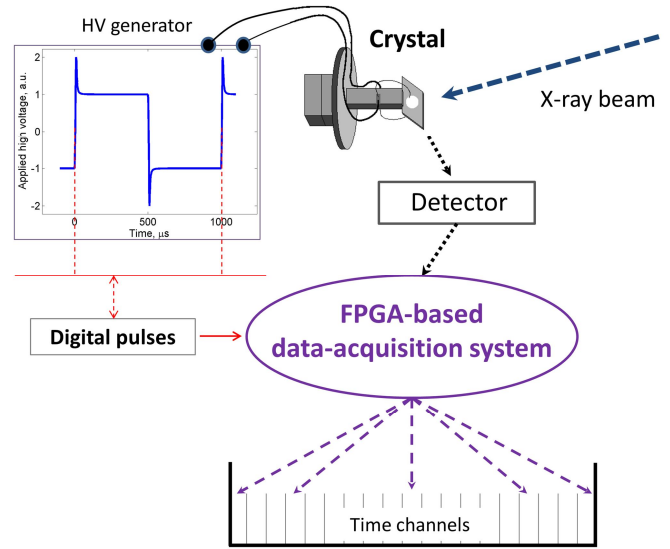

Figure 8 illustrates the rocking curve of the 3 9 2 reflection from the piezoelectric BiB$_3$O$_6$ crystal stored successively into 10000 time channels, each 100 ns long and within a single high-voltage period. The first and 5001st channels are triggered to the positive and the negative high voltage edge (the beginning and the middle of a high-voltage cycle), correspondingly. The jump of the average peak position in the middle of a cycle results from the switching of field polarity. In addition the switch of applied electric field generates the oscillations of the peak position, which can be related to the oscillations of macroscopic strain of the crystal. The oscillations result from the elastic deformation propagating back and force in the crystal; the frequencies of the oscillations (eigen-frequencies of the crystal plate) depend on the geometry of the crystal and its elastic constants / sound velocities.

Figure 8. The timeline of the 3 9 2 Bragg peak rocking curve of the BiB$_3$O$_6$ piezoelectric crystal within a single high-voltage period; the measurements were performed with the data acquisition system shown in Figure 6. The discontinuity appearing at the 5001st time channel corresponds to the change in the HV polarity – at this point the average position of the peak maximum displaces accordingly and oscillates around a new position [52].

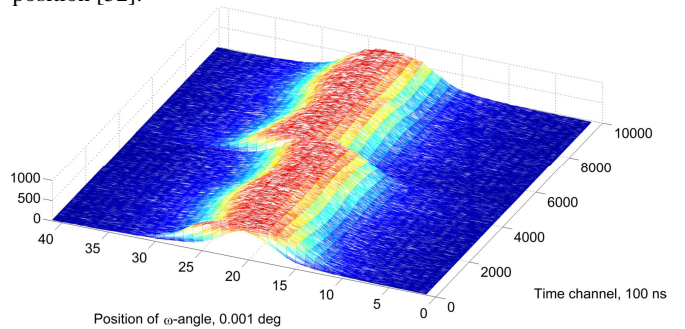

The oscillations, caused by the dynamic switch of the field can be qualitatively described by the model of a generalized harmonic oscillator. The framework of this model consists of a generalized mechanical force *F* as a perturbation and the generalized displacement *q* of a mass *M* as a response. When a mechanical force is applied statically, the response is proportional to the force ($q=F/k$, where *k* is the generalized force constant). In the case of dynamically applied perturbation, the response, *q*, is also dynamic and can be obtained from the equation of motion:

$$M\ddot{q} + 2M\gamma\dot{q} + kq = F(t) . \qquad (6)$$

Here $\gamma$ is the damping constant accounting for the generalized attenuation force. Modelling the time regime of the applied perturbation by the rising time $\tau$ and the expression

$$F(t) = F_0 \left[1 - exp\left\{-\frac{t}{\tau}\right\}\right] \qquad (7)$$

leads to the oscillatory solution of the equation (6). Figure 9 shows the solutions of (6) for two qualitatively different rising times measured in the periods of eigenmode of the oscillator $T_0 = 2\pi\sqrt{k/M}$. The amplitude of the oscillations is significant for a 'fast' ($\tau = 0.1 \cdot T_0$) (Figure 9, left) switch of the perturbation. It diminishes for the 'slow' ($5 \cdot T_0$) (Figure 9, right) switch and disappears when the rising time is increased.

The generalized displacement *q* accounts for either an atomic displacement or a strain. In the first case, the period of free oscillations ranges by the periods of optical phonon vibrations, where the key parameters are interatomic interactions ($T_0 \sim 10^{-12}$ s); electric field dynamics is 'slow'. In the second one, eigenmodes are acoustical phonon vibrations where the key parameters are sound velocities ($T_0 \sim 10^{-6}$ s) ; electric field dynamics is 'fast'. If both the microscopic and the macroscopic deformations were linked to each other, the resulting model oscillator would adapt the longest period of vibration.

Figure 9. The normalized response of a damped harmonic oscillator, $q(t) \cdot k/F_0$ (oscillating curve), to the dynamical switch of external force, $F(t)/F_0$ (smooth curve). Response of a harmonic oscillator is simulated as a solution of an equation of motion (6) for two rising times of the force (7): $\tau = 0.1 \cdot T_0$ (fast switch, left) and $\tau = 5 \cdot T_0$ (slow switch, right), where $T_0$ is the period of eigen-vibrations of the oscillator. The fast switch of the force generates high-amplitude oscillations of the displacements; the slow switch of the force induces lower-amplitude oscillations; further increase of the rising time removes the oscillations: the response will be proportional to the force as in a static case [52].

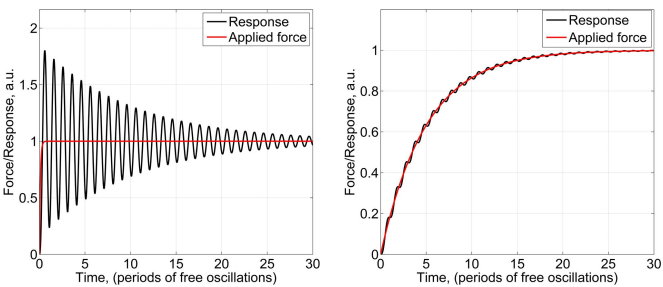

The experiments were performed at the BM01A beamline at the ESRF using the wavelength $\lambda = 0.64$ Å. We collected the rocking curve of the selected Bragg reflections repeatedly; the satisfactory counting statistics was reached after typically 5 hours of the measurement. This data collection time corresponds to approximately $2 \cdot 10^7$ high-voltage cycles applied to the crystal. Figure 10 shows both the shift of the peak position and the relative change in the integrated intensity of the 0 5 1 peak collected from the $Li_2SO_4 \cdot H_2O$ crystal. The time regimes differ substantially: the peak position oscillates around the corresponding static values, so that the equilibrium value of strain is not reached when the direction of electric field is reversed. The integrated intensity does not oscillate and follows the smooth time dependence of the applied external electric field. The maximum observed change of intensity (~0.23%) has the same order of magnitude as the change of intensity, expected under static electric field (0.38%, based on the atomic displacements which were refined using the result of the static experiment). Any modulations in the intensity are smaller than the statistical noise of the experiment. In contrast, the oscillations of the peak position are so strong that the curve passes through zero peak shifts a few times.

Figure 10. The time-dependence of the 0 5 1 peak shift (blue oscillating curve) and the relative change in intensity (red curve) in response to the dynamic switches (~ 3 kV / 200 ns) of an applied electric field. The peak shift oscillates, while the intensity does not; it follows the profile of the applied electric field. This observation agrees with the model of a harmonic oscillator, corresponding to fast and slow switches of applied perturbation.

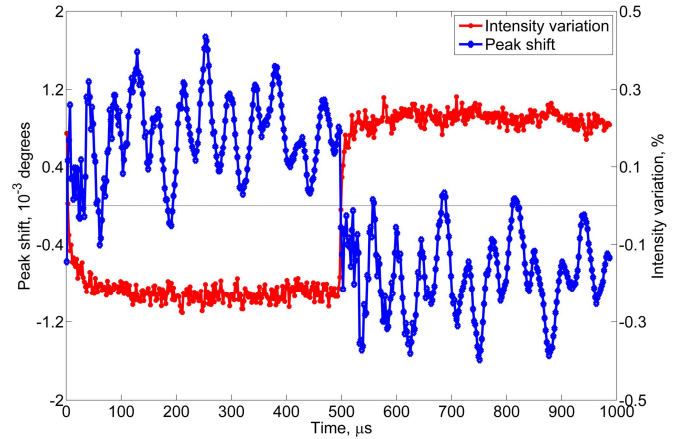

The oscillations of the peak positions and smooth behaviour of the intensity suggest that atomic displacements are the driving force for the piezoelectric deformation. This means that the external electric field displaces the atoms in a unit cell first. The time constant of this process ($T_0 \sim 10^{-12}$) is much smaller than the switching time of the electric field therefore no oscillations are expected (Figure 9, right). The displacements of atoms in a unit cell produce the net internal stress, related to the distortion of the bond lengths and the bond angles not compensated by the external electric field. This internal stress is elastically released by the development of a macroscopic strain, i.e. by piezoelectric deformation. The time constant of this process ($T_0 \sim 10^{-6}$ s) is comparable to the switching time of an applied electric field, therefore the oscillations are expected (Figure 9, left)

The scenario, suggesting that displacements of atoms follow a macroscopic strain must be rejected. If it was the case then the oscillations of the Bragg intensity would have to be observed and correlated with the oscillations of the peak position. The experimental results do not give any clear evidence of such oscillations and show that the diffraction intensity follows the value, close to that expected under a static electric field [25] (~ 0.38 %).

In conclusion, the time resolved experiment suggests that atomic positions in a unit cell are not affected by the macroscopic deformation. Piezoelectric deformation is a result of atomic shifts and distortion of bond lengths/bond angles.

## Conclusion

We have demonstrated that electric field induced bond distortions depend on the bond character, bond strength and atomic charges of the bounded atoms.

The closed shell ionic Li–O interactions have been observed to be the most sensitive to an applied electric field. The distortion of the Li–O bond length is proportional to the projection of the electric field on the bond line. The displacement of the Li atom is almost independent from the displacements of other atoms in the structure. This conclusion agrees well with the relatively weak bonding between Li and O and allows describing Li as an independent body.

The S–O interaction has been shown to be the most resistant to an applied electric field; the distortion of the S–O bond length is 10 times smaller than that of the Li–O bonds in the same crystal. This result is the least understood so far: the small distortion of the S–O chemical bonds contradicts the fact that Bader's atomic charge of sulphur atom is quite big (+4.1 e); it suggests that the electric field force on a sulphur atom should not be described in terms of the point charge model.

The intermediate Ga–O and P–O interactions showed a strong role of the bond polarity (ionicity). The stronger P–O bonds turned out to be more sensitive to electric field than the weaker Ga–O bonds. This difference contradicts the expected bond strengths but agrees with the differences in the atomic charges: it is significantly bigger for the P atoms than for the Ga atoms.

To summarize, our work provides the first proof of principle. It does not present a comprehensive review of bond sensitivity to an applied electric field. More structures containing different types of atomic interactions must be investigated. The number of the structures that can be studied under electric field is, however, limited due to the experimental difficulties. Moreover, the collection of data is very time consuming. The number of improvements in experimental technique is expected: the development of the fast area detectors must facilitate faster data collection.

We verified the mechanism of macroscopic piezoelectric distortions: displacements of atoms in a unit cell are the primary reason for macroscopic deformations. Both processes are separated in time: atomic shifts are completed much faster than macroscopic strain. We make this conclusion on the basis of simultaneous observation of Bragg intensity and peak position in response to a fast (200 ns) switch of electric field polarity.

## Acknowledgement

We acknowledge the financial support of Deutsche Forschungsgemeinschaft within the special priority program SPP 1178 (Experimental charge densities as a tool for understanding intermolecular interactions). VGT thanks Alexander von Humboldt Foundation for the Humboldt Research Award. The members of the Swiss-Norwegian beamline at ESRF, Dr. Dmitry Chernyshov and Dr. Philip Pattison are greatly acknowledged for their experimental support. Dr. Wolfgang Morgenroth and Dr. Martin Tolkiehn are acknowledged for their support at the HASYLAB D3 beamline. We acknowledge Dr. Ladislav Bohatý and Dr. Petra Becker Institute of Crystallography, University of Cologne for providing single crystals of $BiB_3O_6$ and $Li_2SO_4 \cdot H_2O$.